\documentclass[a4paper,11pt]{article}
\usepackage{graphicx}
\begin{document}
\title{Bethe-Schwinger Effective Range Theory and Lehmann and Weinberg Chiral Perturbation Theories}
\author{Tran N. Truong \\
\small \em Centre de Physique Th{\'e}orique,\\
 \small \em Ecole
Polytechnique, Palaiseau, France \\}

 \maketitle

 \centerline{Dedicated to the Memory of Prof. K. Nishijima}

  \centerline {(Paper presented at  ICFP 09, September 24-30, 2009, Hanoi, Vietnam)}

 \vskip 1cm

\begin{abstract}

This paper is a brief review of  low energy soft hadronic physics,
starting from the invention of the low energy effective range
theory in the late 40's due to Bethe and Schwinger for
nucleon-nucleon scattering, and its generalization to
 the static Chew-Low model for pion nucleon
scattering, to the present development of the Lehmann and Weinberg
Chiral Perturbation Theories. It is pointed out that a consistent
low energy calculation can be achieved with the incorporation of
the unitarity relation in the Chiral Perturbation Theory.

\end{abstract}

\section {Introduction}

It is appropriate to write a brief review of the low energy non
relativistic effective range theory as it was first started out in
1948 \cite{schwinger, blatt, bethe}  and the recent development of
the Chiral Perturbation Theory by Weinberg and others (ChPT)\cite{
weinberg1}, the Unitarized Chiral Perturbation Theory by
Lehmann(UChPT)\cite{lehmann} and others \cite{truong1, truong2},
i.e the Chiral Perturbation Theory (ChPT) for pions with the
incorporation of unitarity relation. I want to argue that the
UChPT  is a logical follow-up of the non relativistic effective
range theory for nucleon-nucleon scattering developed in the late
40's by  Bethe  and Schwinger  and  10 years later, the Chew-Low
theory for the $\pi N$ scattering,
 $\Delta$ resonance.
\cite{chew1}.

In fact the development of UChPT by Lehmann \cite{lehmann} was
done 7 years earlier than the perturbation approach of Weinberg
\cite{ weinberg1}. Unfortunately, because Lehmann calculation was
done
 in the chiral limit and that his unitarisation of the partial wave
 amplitudes for the pion pion scattering was done in the old-fashioned
 effective range theory,
  his line of approach to the chiral theories was not appreciated
by workers in ChPT.

The important question is whether one can use perturbation theory
for strong interaction physics? On one hand we have the effective
range theory which implies that strong interaction cannot be
treated perturbatively, on the other hand it is now a fashion to
treat chiral theories by the perturbation theory (ChPT) as
advocated by Weinberg \cite{ weinberg1} and his followers. The
question is which line of approach is correct or are they both
correct?

ChPT may be valid at a very low energy where the constraint of the
elastic unitarity could be unimportant (see, however, the
discussion in the section on the form factor calculation), this
situation is no longer satisfactory as the energy region of
interest is closer to the resonance region. How close or far from
the resonant region is difficult to define precisely. It is
therefore useful to have a theory which is also valid at low
energy and also in the resonance region. As we shall see, the
failure of the ChPT approach in calculating of the phases of the
pion, pion Kaon scalar and vector form factors, in its early
development stage at \textit{one loop level},
 reflects the lack of the
unitarity in the theory \cite{truong1}. Recent ChPT calculations
of these processes at two loop level, in my opinion, have removed
to some extend these difficulties.  It is regrettable, however,
that in recent ChPT calculations  the \textit{phases} of the
matrix elements are not calculated in order to compare with
experiments or ChPT theoretical prescription. For example for $\pi
\pi$ elastic scattering, the phase shifts are identified with the
real part of the amplitudes which is in itself the unitarization
prescription in ChPT \cite{truong1, truong6}; it can also,
however, also
 be identified
with the ratio of the real to imaginary parts of the partial wave
matrix element. The difference of these two calculations reflects
the accuracy of the ChPT approach.

This review is rather partial, emphasizing mostly only on the
dispersive approach where unitarity is respected  which I have
made a number of contributions and which I am familiar with. I
must admit that I am not familiar with most recent enormous ChPT
works and hence I have to concentrate on my previous publications.
Most of the discussions are given by older calculations which I
have not time to update with new experimental results. I think it
is useful to summarize some these old calculations before they got
lost because the current fashion of using perturbation calculation
for strong interaction, although this may be incorrect. Whenever
possible, I shall compare the results of the non-perturbation
approach with those of ChPT. I have to apologize to many authors
whose related works are not discussed in this review.

Being a physicist belonging to the older generation and being
brought up during the early day of the development of strong
interaction particle physics, I was deeply influenced by such
general principles as the unitarity  and dispersion relation
because they were extensively used at that time.

During my second year in the Graduate School at Cornell, I had the
privilege to follow the nuclear physics course taught by Professor
Bethe who lectured  on his low energy effective range theory
\cite{bethe} and also to attend the Particle Physics course where
he lectured on the static Chew-Low effective range theory of the
$\Delta$ resonance \cite{chew1}and also on the application of
dispersion relations. Most physicists nowadays would dismiss these
topics are no longer of interest, but I think quite contrary.

My involvement with chiral symmetry did not come until the late
70's after the discovery of the $\tau$ lepton. My collaboration
with Pham and Roiesnel resultsed in the publication of the current
algebra calculations for the hadronic decays of the $\tau$ lepton
with a follow-up calculation of $e^+ e^- \rightarrow 4\pi$ where
the discrepancy of a factor 20 in the cross section between the
soft pion theorem was explained \cite{pham1, pham2, pham3}. This
result was  independently rediscovered 24 years later by Ecker and
Unterdorfer \cite{ecker}. After these works I became interested in
the $Ke_4$ problem and pointed out the role of analyticity and
unitarity, in particular the role of the threshold square root
singularity for the S-wave pion pion scattering \cite{truong20}.
(It is regrettable that this non-relativistic quantum mechanics
name of the square root singularity is nowadays replaced by the
name of chiral logarithm  which has the threshold square root
singularity!). Following this work, using the idea of the square
root threshold singularity and with the collaboration of C.
Roiesnel, we gave a resolution to the $\eta \rightarrow 3\pi$ rate
which was previously calculated by Weinberg giving a too low rate
by a factor of 5 \cite{ truong21, roiesnel}.

Most of my works on Chiral Symmetry, unlike in the standard ChPT
approach \cite{ weinberg1}, were based on Current Algebra and on
the supposition that the chiral power series Effective Lagrangian
is an effective theory which contains all features of Chiral
Symmetry and Current Algebra; the calculated matrix element is
valid whenever a power series expansion in momenta of the
Nambu-Goldstone boson  is legitimate i.e in the region where there
are no singularities , e.g. no unitarity cut, and is usually in
the unphysical region of energy. The calculated S-Matrix element
from the chiral effective lagrangian has to be analytically
continued to the physical region (on the cut) with the constraint
of  unitarity and analyticity. Because we are interested in a low
energy theory in  the elastic region, the elastic unitarity
relation has to be imposed using its full form and not the
perturbation unitarity relation as usually done in ChPT
\cite{truong21}.

Fortunately in this approach, the elastic unitarity relation
enables us to generate the low energy resonance $\rho$, K* ...
which are the main features of the low energy of the soft pion and
kaon physics. The elastic unitarity relation can be implemented in
the dispersion relation approach by the inverse amplitude method
(IAM), the N/D method or simply the resummation of the
perturbation series by the Pade aproximant method (valid also for
the multichannel problems). This last approach may be a good
compromise for those who love perturbation theory and have
neglected the unitarization procedure in their perturbation
results. Most of my works on $K_{l4}$, pion, $K\pi$ form factors,
hadronic and rare K decays, hadronic $\tau$ decays and the $\eta
\rightarrow 3\pi$ which was done either by myself or with my
collaborators followed this line of approach; the calculation can
be done on a few page of papers, if not on the back of an envelope
as compared with the enormous length of the ChPT calculations
\cite{truong21}.

This viewpoint of the effective lagrangian is similar to the
problem dealt previously in the literature on the question of
deducing physical consequences in the time like region of the form
factor (on the cut)from a knowledge of a few terms of a Taylor's
series expansion in the momentum transfer of the space like form
factors (below the cut)\cite{levinger}. There is of course no
satisfactory answer to this problem. For the  pion form factor at
low energy, below the inelastic region, our answer is that the
elastic unitarity via dispersion relation must be imposed and not
the the technique of conformal mapping of mapping the form factor
cut plane into a unit circle to do the analytic continuation.

 In
related processes, the constraint of the elastic relations forces
us to make use of the solution of the Muskelishvilli-Omnes
\cite{mo} integral equation, the inverse amplitude method, the N/D
method and also the Pade approximant method if the perturbation
method is used. I shall show that the pure perturbation theory is
not applicable in the presence of the $\rho$ resonance for the
vector pion form factor. Although the elastic unitarity constraint
can generate the $\rho$ resonance, it is however not sufficiently
accurate to explain the experimental data, and hence we are forced
to introduce
   additional parameters to
simulate the inelastic effects at low energy. This is so because
in the dispersive approach, the imaginary part of the form factor
or scattering amplitudes, gets contribution from of all higher
mass intermediate states, a satisfactory  low energy theory  must
minimize their contribution as they are difficult to calculate.

For this reason we can also criticize the calculation of the
Chew-Low model in the sense that  the approximation of the elastic
unitarity is made here without introducing a subtraction constant
in order to suppress the contribution from higher mass
intermediate states or the inelastic effect. In the world of
Chiral Symmetry, similarly to the quantum electrodynamics
phenomena (e.g. low energy theorems for Compton scattering), there
are also low energy theorems which can set the scale for our
calculation. The complication in QCD is that the pseudoscalars
$\pi,K, \eta$, unlike the photon have finite mass unlike the
photon.

The approach of analyticity and elastic unitarity for low energy
physics can be criticized for being unsystematic. This may be
true, but let us point out also that although most of the matrix
elements of quantum electrodynamics can undeniably be treated by
the perturbation theory except the bound state problems which must
be treated by the  approach of the Bethe-Salpeter equation of the
ladder summation. Here in UChPT, the constraints of analyticity
and elastic unitarity on the matrix elements, are treated by
 the IAM, the N/D, the Pade approximant methods or the solution
  of the  Muskelishvilli-Omnes
integral equation \cite{mo}.

The plan of this talk is organized as follows:

The first sections is devoted to the explanation of the effective
range theory, which, in the Bethe's approach, is related to the
strength and finite range of the potential in the Schrodinger
equation. It is shown here that in fact the effective range
expansion is due to a more general principle of analyticity and
elastic unitarity for the partial wave amplitudes. This means that
the usual perturbation expansion calculation for the partial wave
amplitudes cannot be consistent with unitarity unless some
summation methods have to be used or that the strength of the
interaction is sufficiently weak. The presence of the low energy
resonances in the pion and Kaon systems (e.g. $\rho, K^*$...)
invalidates this possibility.

The following four sections deal with various unitarisation
schemes.

A brief review of the pion pion scattering is given in  section 6.

 A somewhat detailed study of the vector pion form factor is
given in section 7 in order to explain the difficulties of the
ChPT approach. We show that there are some problems associated
with calculating the pion form factor phase using perturbation
theory as previously discussed in reference \cite{truong1}. We
give here the answer to the question of how to incorporate the
ChPT calculation to the vector meson dominance (we cannot
unambiguously do unless
 the effective lagrangian is used as advocated in our approach).
 Two possibilities
could be tried: a) The most popular one is to use the ChPT result
at some low energy as a low energy theorem to set the scale (as
subtraction constants) for the dispersion relation, b) Another
possibility is to add Vector Meson dominance or dispersion
relation amplitudes to ChPT \textit{amplitudes} at a given order
not just at a few points.

 In either possibility, we run against an amusing
"theorem" stating "there is no such a thing as  a small analytic
function"\cite{dicke}, that is a small unmeasurable analytic
function at low energy can become very large at a higher energy.
For example, a small difference at low energy  between a resonance
amplitude given by the elastic unitarity calculation and that of
ChPT calculation can become enormous in the $\rho$ region.

 A criterium is tentatively given to test under what
circumstances the standard perturbation theory can be used.

In section 8, the $Ke_4$ problem is discussed.

In section 9, the $\eta \rightarrow 3\pi$ is briefly summarized.

In section 10, the $K \rightarrow \pi$, $K \rightarrow 2\pi$, $K
\rightarrow 3\pi$ amplitudes are discussed.

In section 11 the $\gamma \gamma \rightarrow 2\pi$, $K_S
\rightarrow 2\gamma$ and $K_L \rightarrow \pi^0\gamma\gamma $ are
discussed.

In section 12, the $\tau \rightarrow K \pi \nu$ and $\tau
\rightarrow 3\pi \nu$ Decays are mentioned.

Finally in section 13, I briefly discuss the problem related to
the calculation of the \textit{absolute} enhancement factor due to
the final state pion pion interaction in the $K \rightarrow 2\pi$
decay which is of current interest.

\section {Non Relativistic Effective Range Theory and Inverse Amplitude Method }

The history of the development of the Effective Range Theory is a
long one. As early as 1939, Breit and collaborators \cite{breit}
suspected that low energy experiments on nucleon nucleon
scattering can determine only two parameters in  nucleon-nucleon
potential, the effective potential depth and range. Subsequently,
Landau and Smorodinsky \cite{landau, smo}    suggested that an
effective range expansion for the the S-wave phase shifts $\delta$
:
\begin{equation}
k\cot\delta(k^2)=-\frac{1}{a} + \frac{1}{2}r_0k^2+ ...
\label{eq:er}
\end{equation}
where k the relative momenta, a  the scattering length  and $r_0$
the effective range. The minus in front of $a$ is  by convention.
We omit the superscripts for the singlet and triplet states for
convenience. For the triplet  state, $a$ is positive because of
the deuteron bound state, and $a$ is negative for the singlet
scattering.

Schwinger \cite{schwinger, blatt}
  was the first person to give a general
proof of the effective range expansion Eq. (\ref{eq:er}). His
proof is quite complicated and was based on a variational
principle. A year later Bethe \cite{bethe} and others
\cite{peierls, chew2} gave a much simpler proof.

The low energy nucleon-nucleon experimental data on the S-waves
 singlet and triplet states agree very well with the effective
 range expansion,  Eq. (\ref{eq:er}).

 The proof of  Eq.(\ref{eq:er}) using the Shroedinger Equation
 depends only
 on the assumption of the finite range of the potential and not on
 the strength of the potential. It holds for potentials which
  are strong enough to produce
 a resonant  virtual bound state as in the singlet scattering,
  or the real triplet (deuteron) bound
 state. It also holds for a  weak  scattering potential with a finite range.

 Bethe's proof is based on the following physical picture. Let us
 divide the spatial scattering region into two separate regions,
 inside and outside the potential. Outside the potential range,
 the scattering wave function is that of the asymptotic form with
 the shifted phase shifts, inside the potential, the wave function
 is distorted under the influence of the action of the potential.
 The scattering length $a$ is the zero energy wave function which
 intercepts on the distance axis, and the effective range is
 proportional to the integral of the difference of the square of
 the true and asymptotic wave function. Because one works with the
 Schroedinger equation, it is expected that the effective range expansion
 is consistent with unitarity.

 In order to generalize Eq. (\ref{eq:er}), to a relativistic
 situation which is now the central point of the development of the low energy pion physics,
  in particular the ChPT, it is useful to
 derive it from the more general principles of analyticity and
 unitarity of the partial without referring explicitly to the type of potentials,
 except that they are of finite range.

 Let us consider, for example, the S-wave scattering amplitude $f(\nu)$ and omit the
 subscripts or superscripts spin and isospin. Setting $\nu=k^2$, the elastic
 unitarity  relation is:
 \begin{equation}
Imf(\nu) = \rho(\nu)\mid f(\nu) \mid ^2  \label{eq:un}
\end{equation}
where  $\rho(\nu)= \sqrt{\nu}$ is the non-relativistic phase space
factor and $\nu$ is the square of the relative momentum $k$. Eq.
(\ref{eq:un}) implies that:
\begin{equation}
f(\nu) = \frac{e^{i\delta(\nu)}\sin\delta(\nu)}{\rho(\nu)}
\label{eq:ph1}
\end{equation}
which is the same as:
\begin{equation}
f(\nu) = \frac{1}{\rho(\nu)(-i+\cot\delta(\nu))} \label{eq:ph2}
\end{equation}
and hence any function  representing $\delta$, in particular, for
$tan\delta$ or $cot\delta$  used in Eq. (\ref{eq:ph1}) or Eq.
(\ref{eq:ph2}) would give rise to a  partial wave amplitude
satisfying the unitarity relation.

There is however a restriction on the choice of the appropriate
function, namely the analytic property of the constructed partial
wave amplitudes which can be proved from general principles
\cite{truong3}. The partial waves are, in fact, analytic functions
in the complex $\nu$-plane with a right cut on the real axis from
$0$ to $\infty$ and a left cut from
 $-\nu_c$ to $-\infty$ where $\nu_c$ is positive. On the right hand cut the unitarity
relation Eq. (\ref{eq:un}) must be satisfied and hence this cut is
usually referred as the unitarity cut. The discontinuity across
the left cut depends on the characteristic of the potential used
in the Schroedinger equation to describe the scattering process.
The analytic  and unitarity properties of the partial waves are
well explained in an article by Blankenbecler et. al.
\cite{blank}.

Let us define the inverse function $g(\nu) = f^{-1}(\nu)$. This
function is also analytic with the same right and left cuts, apart
from isolated poles coming from the zeroes of $f(\nu)$. On the
positive real axis $g(\nu)$ is given by:
\begin{equation}
g(\nu) = \rho(\nu)(-i+\cot\delta(\nu)) \label{eq:inv}
\end{equation}
and it can  analytically be continued throughout the complex
$\nu$-plane, because its singularity are just on the real axis and
isolated poles.

Let us now write an once subtracted dispersion relation  for
$g(\nu)$ using the subtraction point at $\nu=0$. (More than one
subtraction at $\nu=0$ was not possible because the dispersion
integral would diverge; more than one subtraction could, however,
be made if the subtraction point  $\nu_0$ was taken in the gap
$-\nu_c< \nu_0 <0$ ). For simplicity, we assume that $f(\nu)$ does
not have any zero in the complex $\nu$ plane:
\begin{equation}
g(\nu) = g(0) -\frac{\nu}{\pi} \int_{0}^\infty
dz\frac{\sqrt{z}}{z(z-\nu-i\epsilon)} + \frac{\nu}{\pi}
\int_{-\infty}^{-\nu_c} dz\frac{Img(z)}{z(z-\nu)} \label{eq:drg}
\end{equation}
where we use for $\nu>0$, $Img(\nu)=-\sqrt{\nu}$ and for
$\nu<-\nu_c$, $Img(\nu)=-\mid g(\nu)\mid^2 Imf(\nu)$. Once we know
$Imf(\nu)$ on the left cut, e.g. by perturbation theory, we arrive
at a non linear integral equation for $g(\nu)$, just the same as
in the Chew Low theory.

 The first
integral on the R.H.S of this equation can readily be evaluated by
considering $i\sqrt{\nu}$ as an analytic function in the cut plane
with a cut on the real axis from $0$ to $\infty$. Separating the
contribution from the first integral into the principal part and
the $\delta$-function contributions, we finally arrive at:
\begin{equation}
g(\nu) = g(0) + L(\nu) -i\sqrt{\nu}  \label{eq:gn}
\end{equation}
where $L(\nu)$ denotes the second integral on the R.H.S. of
Eq.(\ref{eq:drg}) which is the left  cut contribution i.e. the
potential contribution. Instead of solving the integral equation,
for the present purpose, let us treat it phenomenologically.  For
$\nu>0$, we can expand $L(\nu)$ in a power series of $\nu$,
$L(\nu)= \nu \sum_{n=0}^{\infty} \alpha_n\nu^n$ with a radius of
convergence $\nu_c$. In the special case when a first few terms
are sufficient, from Eq.(\ref{eq:gn}) and Eq.(\ref{eq:inv}), one
has:
\begin{equation}
\sqrt{\nu}\cot\delta(\nu) = g(0) + \alpha_0\nu + \alpha_1\nu^2 +
... \label{eq:efr}
\end{equation}
and is just the effective range expansion of Eq.(\ref{eq:er}). The
scattering length is inversely proportional to the subtraction
constant $-a^{-1}=g(0)$ ,  the effective range is the first term
in the power series; the next term $\alpha_1$ is the potential
shape dependence term.

The expansion of $L(\nu)$ as a power series in $\nu$ has usually a
small radius of convergence. It is better to approximate this
series by a Pad{\'e} approximant method i.e. the ratio of two
polynomials \cite{ba, zin}. The zeroes of the polynomial in the
denominator become poles in the Pad{\'e} approximant. If they were
on the the real negative $\nu$ axis, as they should be, we would
have the usual pole approximation for the left hand cut.  In the
the full relativistic theory for $\pi\pi$ scattering, to be
discussed later, we shall treat the left hand cut contribution
more realistically.

The above treatment is valid for the S-wave nucleon-nucleon in the
singlet state. For the triplet S-wave  nucleon-nucleon scattering,
because of the deuteron bound state, our treatment must be
modified to include the deuteron bound state pole. We then obtain
the same effective range expansion, but the two parameters
scattering length and effective range, are directly related to the
binding energy of the deuteron and the residue of the deuteron
pole.

What has been discussed previously is not new. In fact it was
Noyes and Wong \cite{nw} who recognized first that the effective
range expansion is due to analyticity and unitarity with the
contribution from the left hand cut is apprroximated by a pole
approximation. The pole approximation can be regarded as the Pade
approximant  for the low energy subtraction constants as discussed
above.

It may happen that the partial wave amplitudes obtained from the
inverse amplitude method (I.A.M) could develop poles on the
negative cuts and in the complex $\nu$ plane. They must be removed
from the constructed amplitude. This removal would result in a
violation of the unitarity relation. If the unwanted poles are far
from the threshold region with small residues, their effects on
the violation of the unitarity would be small in the low energy
region and hence could be neglected.

The IAM method, being the simplest one, is not the only method to
unitarize the partial wave amplitudes. Because the partial waves
have both right and left cuts, they can be written as the product
of two cuts whereas the in the I.A.M., they are written as the sum
of two cuts. This is the $N/D$ method. In this method, just the
same as in the I.A.M.,  the right cut is treated exactly, and the
left cut is treated approximately.

Another method which can be quite useful is the Pad{\'e}
approximant method (P.A.M.) \cite{ba, zin}. In this method, the
elastic unitarity relation is satisfied, and the left cut is
treated perturbatively. This method is particularly useful for
perturbation calculations, since the reconstructed series satisfy
unitarity and analyticity. We shall come back to the P.A.M. method
later when we discuss $\pi\pi$ scattering.

The non relativistic K-matrix approach has to be modified in order
to take into account of the analytic properties of the partial
wave amplitudes, i.e. the real and imaginary part of the partial
waves are related to each other by dispersion relation or Hilbert
transform \cite{truong3}.

\section { Relativistic Effective Range Theory}

For a relativistic theory such as $\pi\pi$ scattering, the
phenomenological approach to the effective range theory can
similarly be carried out. The phase space factor is now
$\rho(\nu)=\sqrt{\frac{\nu}{\nu+\mu^2}}$ where $\mu$ is the pion
mass. The partial wave amplitude $f(\nu)$ is defined as in Eq.
(\ref{eq:ph1}), with the new expression for the phase space
factor. $f(\nu)$ has the same analytic structure as the
non-relativistic one:
\begin{equation}
f(\nu) = \frac{e^{i\delta(\nu)}\sin\delta(\nu)}{\rho(\nu)}
\label{eq:phr}
\end{equation}
For simplicity let us consider from now on, as an example, the
S-wave amplitude; higher partial waves because of the kinematical
zeros at threshold can be straightforwardly taken care of.
Following the same reasoning as in the non-relativistic situation
with the same definition for the inverse amplitude
$g(\nu)=f^{-1}(\nu)$, the contribution of the principal part of
the first integral on the R.H.S. no longer vanishes. Instead of
Eq.(\ref{eq:efr})  we arrive at the relativistic effective range:
\begin{equation}
\rho(\nu)\cot(\delta) = g(0) +
\frac{2}{\pi}\sqrt{\frac{\nu}{\nu+\mu^2}}\ln\frac{\sqrt{\nu}+\sqrt{\nu+\mu^2}}{2\mu}+
\alpha_0\nu + \alpha_1\nu^2 + ... \label{eq:efrr}
\end{equation}
with $g(0)=\mu/a$, where $a$ is the scattering length and we have
used the sign convention for the relativistic scattering length,
i.e. $a>0$ for an attractive interaction. In addition to
$\alpha_0$ which is proportional to the effective range, we have
the additional logarithm term.

As in the non-relativistic case Eq. (\ref{eq:gn}) , $L(\nu)$ can
be expanded in a power series and it is convenient to use of the
Pad{\'e} approximant method \cite{ba, zin}  for this power series.

A generalization of the effective range expansion for the theory
of the P-wave, isospin $3/2$ $\pi N$ scattering, the $\Delta$
resonance is obtained, using the same line of reasoning. Using
 the dispersion relation for the P-wave for $\pi N$
scattering in isospin 3/2 and taking into account of  the nucleon
poles in the direct and crossed channels and requiring the elastic
unitarity condition, one would get a non linear integral equation
of the Chew-Low theory \cite{chew2}. Using the I.A.M., the
unitarity relation can then be treated exactly, and then the left
cut is to be treated by perturbation as a first iteration.

The development of the effective range Chew-Low theory
\cite{chew2} is therefore simply a generalization of the
non-relativistic theory.
 This  is a first triumph
for using analyticty and elastic unitarity to solve a non
perturbation  problem in a relativistic pion physics.

We shall see below, there is also a simple method which can resum
the perturbative approach in a manner that unitarity is satisfied.
This is the Pad{\'e} approximant method (PAM) \cite{ba, zin,
truong1}.

\section {  Pad{\'e} Approximant Method}

Let us write the partial wave perturbation series as:
\begin{equation}
f^{pert}(\nu) = f^0 + f^1(\nu) +... \label{eq:pt}
\end{equation}
where $f^0$ is the tree amplitude which is assumed here, for
simplicity, a constant or a real polynomial (otherwise it has only
the left hand cut singularity). $f^1(\nu)$ is the one loop
amplitude satisfying the perturbative unitarity, for $\nu>0$:
\begin{equation}
 Imf^1(\nu) = \rho(\nu)(f^0)^2  \label{eq:pu}
\end{equation}
Let us construct the $[0,1]$ Pad{\'e} approximant:
\begin{equation}
f^{[0,1]}(\nu)=\frac{f^0}{1-\frac{f^1(\nu)}{f^0}} \label{eq:pd}
\end{equation}
This equation gives rise to a geometric series constructed out of
$f^0$ and $f^1$. Expanding the denominator of Eq.(\ref{eq:pd}) in
a power series of $f^1/f^0$, its first two terms agree with the
perturbation expansion, Eq.(\ref{eq:pt}). The presence of the
remaining terms is to preserve the elastic unitarity condition
because:
\begin{equation}
Imf^{[0,1]}(\nu)=\rho(\nu)\mid f^{[0,1]}(\nu) \mid^2
\label{eq:upd}
\end{equation}
for $\nu>0$. For $\nu<-\nu_c$, the discontinuity of $f^{[0,1]}$
across the left hand cut is $2i$ times the imaginary part of the
Pad{\'e} amplitude is given by:
\begin{equation}
Imf^{[0,1]}(\nu)=Imf^1(\nu) \frac{\mid f^{[0,1]}(\nu)\mid ^2
}{(f^0)^2} \label{eq:dl}
\end{equation}
and hence the same as the perturbation result but it is modified
by a factor $\mid f^{[0,1]}\mid^2/(f^0)^2$.

The phase shifts given by the Pad{\'e} amplitude is:
\begin{equation}
\rho(\nu)\cot\delta(\nu) =
\frac{1}{f^0}-\frac{Ref^1(\nu)}{(f^0)^2}
 \label{eq:erp}
\end{equation}
and is a generalisation of the relativistic effective range
expansion of Eq.(\ref{eq:efrr}). At some energy, if there was a
cancellation of the two terms on the R.H.S. of this equation, a
resonant state is generated.

 The Pad{\'e} approximant method is similar to the
bubble summation for the partial wave \cite{wil}, but  is more
general, because it has both unitary (right) and left cuts,
whereas the bubble summation has only the unitarity cut.

There are  few  methods in the relativistic particle physics to
treat the non-perturbation problem: the infinite geometric series
of the bubble summation used in the study of the
Nambu-Jona-Lassinio model, or the infinite ladder summation of the
Bethe-Salpeter equation used to treat the bound state problem in
Quantum Electrodynamics. These treatments are not as systematic as
the perturbation series, but they have successfully been used to
treat the non perturbation phenomena.

\section { Relation Between Pad{\'e} Approximant Method (P.A.M) and Inverse
Amplitude Method (I.A.M)}

Within some approximation, the Pad{\'e} approximant method can be
derived from the more general I.A.M..  To see this let us write
down the dispersion for the inverse of the partial wave
$g(\nu)=f^{-1}(\nu)$ and use the elastic unitarity condition:
\begin{equation}
g(\nu) = g(0)-g(\nu) = g(0) -\frac{\nu}{\pi} \int_{0}^\infty
dz\frac{\rho(\nu)}{z(z-\nu-i\epsilon)} - \frac{\nu}{\pi}
\int_{-\infty}^{-\nu_c} dz\frac{\mid g(z) \mid^2 Imf(z)}{z(z-\nu)}
\label{eq:rdrg}
\end{equation}
where it is assumed that there are no  zeroes in $f(\nu)$. Unless
$Imf(\nu)$ is given on the left cut, we cannot proceed. It is
usually assumed that $Imf(\nu)$ is given by the perturbation
series on the left cut and hence Eq.(\ref{eq:rdrg}) becomes a non
linear integral equation for $g(\nu)$. An iteration procedure can
be used to solve this non linear integral equation.

If in the first iteration cycle, one sets on the left cut,
$Imf(\nu)= Imf^{1}(\nu)$ and $\mid g(\nu)\mid^2=(f^0)^{-2}$ then
one would get the Pad{\'e} result, Eq.(\ref{eq:pd}). Hence there
is a closed connection between the P.A.M and the I.A.M. In using
the P.A.M method one thus avoids the problem of solving the
non-linear integral equation as a first approximation.

The $N/D$ method is an attempt to linearize the non-linear
integral equation obtained by the I.A.M.. One write in this case
$f(\nu)=N(\nu)/D(\nu)$ with $N(\nu)$ contains only the left cut
and $D(\nu)$ only the right cut:
\begin{equation}
N(\nu) =
\frac{1}{\pi}\int_{-\infty}^{-\nu_c}dz\frac{ImN(z)}{z-\nu}
 \label{eq:n}
\end{equation}
and:
\begin{equation}
D(\nu) =1 + \frac{\nu}{\pi}
\int_{0}^{\infty}dz\frac{ImD(z)}{z(z-\nu)}
 \label{eq:d}
\end{equation}
with $ImN(\nu)=D(\nu)Imf(\nu)$ and $ImD(\nu)=-\rho(\nu)N(\nu)$.
Using these relations in Eq. (\ref{eq:n}) and in Eq. (\ref{eq:d}),
we arrive at a coupled linear integral equation, instead at a
non-linear one. Its connection with the I.A.M. or P.A.M. is not
obvious. One could try , for example, to lump all functions in the
perturbation series with the left cut singularity with the $N$
function and then use the $N/D$ method to unitarize the
perturbation series.

\section{Low Energy Pion Pion Elastic Scattering }

The S and P-waves pion-pion scattering were first calculated by
Weinberg \cite{weinberg2} using current algebra technique and the
assumption of a power series expansion for the scattering
amplitude.

The effective range for P-wave $\pi\pi$ scattering was first
proposed by Brown and Goble \cite{bg} in the approximation where
the left hand cut contribution was neglected. In a systematics
approach, there was no reason to neglect this contribution because
it is of the same order as the as the logarithm term coming from
the unitarity (right hand cut). Within this approximation Brown
and Goble obtained the Kawarabayashi, Suzuki, Riazuddin and
Fayyazuddin  (KSRF) relation relating the $\rho$ width with its
mass \cite{ksfr}.
 Subsequently,
Lehmann \cite{lehmann}, in order to understand the gross features
of $\pi\pi$ scattering up to  $0.8 GeV$ or so, below which, the
assumption of the elastic unitarity is still valid, did the one
loop ChPT calculation by perturbation theory, but in the limit of
the pions as zero mass Goldstone bosons, using the non-linear
$\sigma$ model ($NL\sigma M$). Because working in the chiral
limit, the results of his calculation depend on only two
parameters. He then unitarized his results by the effective range
expansion in $\cot\delta$ in its simplest form (a power series
expansion in $\nu$), he got scattering amplitudes which are
consistent with analyticity and unitarity. Because his
calculations were done with a zero mass pion, threshold parameters
such as  scattering lengths could not be calculated.

Lehmann  \cite{lehmann}got an overall satisfactory agreement with
the experimental data. The P-wave amplitude, has a reonance $\rho$
at the right mass, and its width, satisfies the KSRF relation
\cite{ksfr}. The I=0 S-wave phase shifts around 0.5 GeV are large
and attractive, the I=2 phase shifts at the same energy are
repulsive and small, in agreement with the experimental data. One
should consider these results as impressive, considering that
there are only two parameters in the calculation.

The lesson from Lehmann calculation \cite{lehmann}, just the same
as the classical calculation of the Chew Low theory \cite{chew1}
and the non relativistic effective range theory, analyticity in
combination of unitarity, enables us to handle the long range
(soft) strong interaction problem, even in the presence of bound
states and resonances.

Lehmann  \cite{lehmann} was puzzled by the P-wave $\pi \pi$
scattering calcualtion of Brown and Goble \cite{bg} where there
was a presence of the logarithm term, whereas in his own
calculation there was no such a term. The answer to this question
is due to the approximation of neglecting the left hand cut
contribution in Brown and Goble calculation. In the chiral limit
the logarithm term from the right and left cut due to the
contribution of the pion loops cancels each other  out to get the
Lehmann result\cite{truong7}. The neglect of the left hand cut
contribution for P-wave, as assumed by Brown and Goble,
 can be justified using the Roy equation
and taking into account of the contribution of the scalar and
vector mesons $\sigma$ and $\rho$ in the t and u channels. The
sign of the crossing matrix is such that their contribution tends
to cancel each other. This is the general justification for the
KSRF relation \cite{truong7}.

A few years later, Jhung and Willey \cite{willey}, improved the
Lehmann $\pi\pi$ scattering calculation using the large $\sigma$
mass limit of the linear-$\sigma$ model, with chiral symmetry
breaking taken into account. They calculated the partial wave
amplitudes to one loop, and then the unitarization procedure was
made by the Pad{\'e} method. A good agreement with experimental
data were obtained, in particular, the $\rho$ resonance.

Lehmann and Jhung and Willey works were done later than the
previous works by Lee and Lee and Basdevant \cite{lee}and others
\cite{zin} based on linear $\sigma$ model and the Pad{\'e}
approximant method
 \cite{zin}. There were not much works thereafter on the low energy
$\pi\pi$ scattering until we publish a paper on the pion form
factors, using the I.A.M. or the P.A.M. methods to take into
account of the unitarity \cite{truong1}. A few years later, a
unitarized version of the one loop ChPT, with chiral symmetry
breaking, was published by Dobado, Herrero and Truong
\cite{truong2}. Excellent agreements with experimental data were
obtained. The  general KSFR relation was recovered here.

The revival of ChPT was due to Weinberg \cite{weinberg1}. He
outlined a systematic perturbation program with chiral symmetry
breaking taken into account. This program has been carried out by
Gasser and Leutwyler \cite{gasser1} and others for scattering
processes and also for the form factor problems \cite{gasser2}.
ChPT for $\pi\pi$ scattering was later carried out even to two
loops order which requires considerable amount of effort
\cite{knecht, bij}. The main emphasis in these calculations is the
systematics approach of the ChPT which is undisputable. However
the crucial point for strongly interacting physics is the
unitarity constraint which was left untouched. Probably, in
analogy with the calculations in Quantum Electrodynamics, it has
been assumed that perturbation unitarity is sufficient. Our
discussion above shows that this is not so.

As far as the $\pi\pi$ scattering is concerned, the perturbative
approach, which naturally led to an expansion in essentially a
power series of $\tan\delta$, is not on the right direction to
treat the non perturbation effect such as the resonant scattering
as explained in section 7.  The relevant expansion should  be an
expansion for $\cot\delta$ as a power series of energy as
explained in section 7. ( For this reason, Lehmann \cite{lehmann},
in the same line of approach as the non-relativistic effective
range theory and the Chew Low theories, was successful in
calculating the $\rho$ resonance). Gasser and Leutwyler
\cite{gasser1} and others, could not get the $\rho$ resonance.

Recent important works incorporating unitarity and analyticity by
the inverse amplitude and the Pad{\'e} approximant methods was
done by Hannah \cite{han} who made a careful study of various
methods and compared them together. The problems of the chiral
zeroes, which we ignored until now because of its complications,
could be straightforwardly taken into account and were well
treated by Hannah. For readers who wished to understand this
subject better, the papers published by Hannah or his thesis could
be quite useful \cite{han, pelaez, oller}.

More recent analysis of the $\pi\pi$ scattering problem in ChPT
was done together with the use of the Roy Equation \cite{roy} and
very good predictions on the S-waves scattering
lengths\cite{colangelo} were obtained. The question is whether
this approach can be simplified and be carried out with the
interpretation of the effective lagrangian as proposed in the
introduction.

\section{Vector Pion Form Factor
Calculation}

 The standard
procedure of testing  ChPT calculation of the pion form factor
\cite{gasser3}, which claims to support the perturbative scheme,
is shown here to be unsatisfactory. This is so because the
calculable terms are extremely small, less than 1.5\% of the
uncalculable terms at an energy of 0.5 GeV or lower whereas the
experimental errors are of the order 10-15\%.

Although dispersion relation (or causality) has been tested to a
great accuracy in the forward pion nucleon and nucleon nucleon  or
anti-nucleon scatterings at low and high energy, there is no such
a test for the form factors. This problem is easy to understand.
In the former case, using  unitarity of the S-matrix, one
rigourously obtained the optical theorem relating the imaginary
part of the forward elastic amplitude
 to the total cross section which is a measurable quantity. This result together with
dispersion relation  establish a general relation between the real
and imaginary parts of the forward amplitude
\cite{truong1,truong17,truong18}.

There is no such a rigourous relation, valid to all energy, for
the form factor. In low energy region, the unitarity of the
S-matrix in the elastic region gives a relation between the phase
of the form factor and the P-wave pion pion phase shift, namely
they are the same \cite{watson}. Strictly speaking, this region is
extended from the two pion threshold to $16m_\pi^2$ where the
inelastic effect is rigourously absent. In practice, the region of
the validity of the phase theorem can be extended to 1.1-1.3 GeV
because  the inelastic effect is negligible.  Hence, using the
measurements of the modulus of the form factor and the P-wave
phase shifts,  both the real and imaginary parts of the form
factors are known. Beyond this energy, the imaginary part is not
known. Fortunately for the present purpose of testing of locality
(dispersion relation) and of the validity of the perturbation
theory at low energy, thanks to the use of subtracted dispersion
relations, the knowledge of the imaginary part of the form factor
beyond 1.3 GeV is unimportant.

Because the vector pion form factor $V(s)$ is an analytic function
with a cut from $4m_\pi^2$ to $\infty$, the $n^{th}$ times
subtracted dispersion relation for $V(s)$ reads:
\begin{equation}
V(s)=a_0+a_1s+...a_{n-1}s^{n-1}+
\frac{s^{n}}{\pi}\int_{4m_\pi^2}^\infty
\frac{ImV(z)dz}{z^{n}(z-s-i\epsilon)} \label{eq:ff1}
\end{equation}
where $n\geq 0$ and, for our purpose, the series around the origin
is considered. Because of the real analytic property of $V(s)$, it
is real below $4m_\pi^2$. By taking the real part of this
equation, $ReV(s)$ is related to the principal part of the
dispersion integral involving the $ImV(s)$ apart from the
subtraction constants $a_n$.

 The
polynomial on the R.H.S. of Eq. (\ref{eq:ff1}) will be referred in
the following as the subtraction constants and the last term on
the R.H.S. as the dispersion integral (DI). The evaluation of DI
as a function of $s$ will be done later.
  Notice that
$a_n=V^n(0)/n!$ is the coefficient of the Taylor series expansion
for $V(s)$, where $V^n(0)$ is the nth derivative of $V(s)$
evaluated at the origin. The condition for  Eq. (\ref{eq:ff1}) to
be valid was  that, on the real positive s axis, the limit
$s^{-n}V(s)\rightarrow 0$ as $s\rightarrow \infty$. By the
Phragmen Lindeloff theorem, this limit would also be true in any
direction in the complex s-plane  and hence it is straightforward
to prove Eq. (\ref{eq:ff1}). The coefficient $a_{n+m}$ of the
Taylor's series  is given by:
\begin{equation}
a_{n+m} = \frac{1}{\pi}\int_{4m_\pi^2}^\infty
\frac{ImV(z)dz}{z^{(n+m+1)}}\label{eq:an}
\end{equation}
where $m\geq 0$. The meaning of this equation is clear: under the
above stated assumption, not only the coefficient $a_n$ can be
calculated but all other coefficients $a_{n+m}$ can also be
calculated. The larger the value of $m$, the more sensitive is the
value of $a_{n+m}$ to the low energy values of $ImV(s)$. In
theoretical work such as in ChPT approach, to be discussed later,
the number of subtraction is such that to make the dispersion
integral converges.

The elastic unitarity relation for the pion form factor is $
ImV(s)= V(s)e^{-i\delta(s)}sin\delta(s) $ where $\delta(s)$ is the
elastic P-wave pion pion phase shifts. Below the inelastic
threshold of $16m_\pi^2$ where $m_\pi$ is the pion mass,
 $V(s)$ must have the phase of $\delta(s)$ \cite{watson}. It is an
experimental fact that  below $1.3 GeV$ the inelastic effect is
very small, hence, to a good approximation, the phase  of $V(s)$
is  $\delta$ below this energy scale.

\begin{equation}
ImV(z)
                        =\mid V(z)\mid\sin\delta(z) \label{eq:ieu}
\end{equation}
and
\begin{equation}
ReV(z)
                        =\mid V(z) \mid\cos\delta(z) \label{eq:reu}
\end{equation}

where $\delta$ is the strong elastic P-wave $\pi\pi$ phase shifts.
Because the real and imaginary parts are related by  dispersion
relation, it is important to know accurately $ImV(z)$ over a large
energy region. Below 1.3 GeV, $ImV(z)$ can be determined
accurately because the modulus of the vector form factor
\cite{barkov,aleph} and the corresponding P-wave $\pi\pi$ phase
shifts are well measured \cite{proto, hyams, martin} except at
very low energy.

It is possible to estimate the high energy contribution of the
dispersion integral by fitting the asymptotic behavior of the form
factor by the expression, $V(s)=-(0.25/s)ln(-s/s_\rho)$ where
$s_\rho$ is the $\rho$ mass squared.

Using Eq. (\ref{eq:ieu}) and Eq. (\ref{eq:reu}), $ImV(z)$ and
$ReV(s)$ are determined directly from experimental data and are
shown, respectively, in Fig.1 and Fig.2.

In the following, for definiteness, one assumes
$s^{-1}V(s)\rightarrow 0$ as $s\rightarrow \infty$ on the cut,
i.e. $V(s)$ does not grow as fast as a linear function of $s$.
This assumption is a very mild one because theoretical models
assume that the form factor vanishes at infinite energy as
$s^{-1}$. In this case, one can write a once subtracted dispersion
relation for $V(s)$, i.e. one sets $a_0=1$ and $n=1$ in Eq.
(\ref{eq:ff1}).

\begin{center}
\begin{figure}
  \includegraphics[width=10cm]{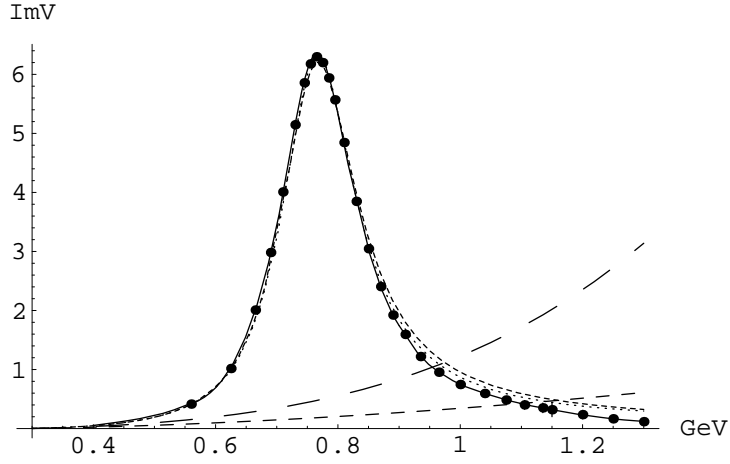}
  \caption{Imaginary Parts of Pion Form Factor ImV as a function of energy in GeV unit.
  The solid curve is a fit to experimental results with experimental errors;
  the long-dashed curve is the two-loop ChPT calculation;
  the medium long-dashed curve is the one-loop ChPT calculation; the short-dashed curve
  is from the one-loop UChPT calculation \cite{truong1} with presumably inelastic
   effects taken into account; the dotted curve is from the UChPT of Hannah }\label{fig:ImV}
\end{figure}
\end{center}


\begin{center}
\begin{figure}
  \includegraphics[width=10cm]{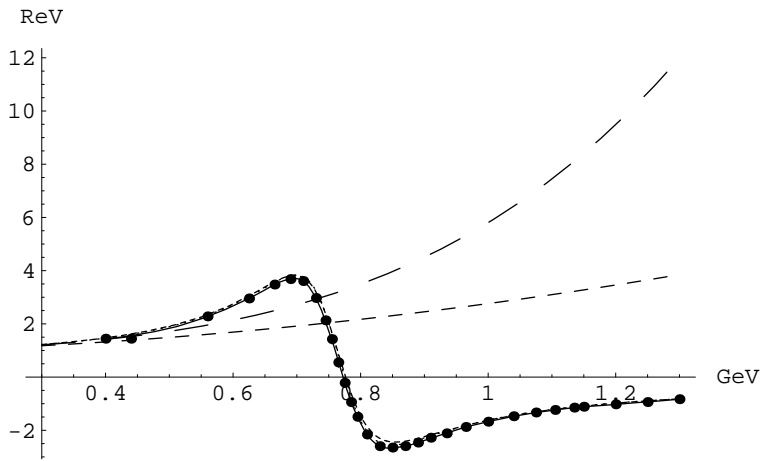}\\
  \caption{Real Parts of Pion Form Factor ReV as a function
  of energy in GeV unit. The label of the curves are the same as
  in Fig. 1; the calculated Real Part of the pion form factor by the once subtracted
  dispersion relation using the experimental imaginary
   part cannot be distinguished from the solid line experimental curve}\label{fig:ReV}
\end{figure}
\end{center}

From this assumption on the asymptotic behavior of the form
factor, the derivatives of the form factor at $s=0$ are given by
Eq. (\ref{eq:an}) with n=1 and m=0. In particular one has:

\begin{equation}
 <r_V^2> = \frac{6}{\pi}\int_{4m_\pi^2}^\infty
\frac{ImV(z)dz}{z^2}\label{eq:rms}
\end{equation}
where  the standard definition $V(s) = 1 + \frac{1}{6} <r_V^2>s+c
s^2 + d s^3+...$ is used. Eq.(\ref{eq:rms}) is a sum rule relating
the pion rms radius to the magnitude of the time like pion form
factor and the P-wave $\pi\pi$ phase shift measurements.  Using
these data, the derivatives of the form factor are evaluated at
the origin:
\begin{equation}
<r_V^2> = 0.45\pm 0.015 fm^2; c = 3.90\pm 0.20 GeV^{-4}; d =
9.70\pm 0.70 GeV^{-6} \label{eq:rvn}
\end{equation}
where the upper limit of the integration is taken to be $1.7
GeV^2$. By fitting $ImV(s)$
 by the above mentioned asymptotic expression, the
contribution beyond this upper limit is completely negligible.
From the 2 $\pi$ threshold to $0.56 GeV$ the experimental data on
the  the phase shifts are either poor or unavailable, an
extrapolation procedure based on some model calculations to be
discussed later, has to be used. Because of the threshold behavior
of the P-wave phase shift, $ImV(s)$ obtained by this extrapolation
procedure is small. They contribute, respectively, 5\%, 15\% and
30\% to the $a_1, a_2$ and $a_3$ sum rules. The results of Eq.
(\ref{eq:rvn}) change little if the $\pi\pi$ phase shifts below
$0.56 GeV$ was extrapolated using an effective range expansion and
the modulus of the form factor using a pole or Breit-Wigner
formula.

The only experimental data on the derivatives of the form factor
at zero momentum transfer  is the root mean square radius of the
pion, $r_V^2= 0.439\pm.008 fm^2$ \cite{na7}. This value is very
much in agreement with that determined from the sum rules. In fact
the sum rule for the root mean squared radius gets overwhelmingly
contribution from the $\rho$ resonance as can be seen from Fig.1.
The success of the calculation  of the r.m.s. radius is a first
indication that causality is respected and also that the
extrapolation procedures to low energy for the P-wave $\pi\pi$
phase shifts and for the modulus of the form factor  are
legitimate.

Dispersion relation for the pion form factor is now shown to be
well verified by the data over a wide energy region. Using
$ImV(z)$ as given by Eq. (\ref{eq:ieu}) together with the once
subtracted dispersion relation, one can calculate the real part of
the form factor $ReV(s)$ in the time-like region and also $V(s)$
in the space like region. Because the space-like behavior of the
form factor is not sensitive to the calculation schemes, it will
not be considered here. The result of this calculation is given in
Fig.2. As it can be seen, dispersion relation results are well
satisfied by the data.

The i-loop ChPT result can be put into the following form, similar
to Eq. (\ref{eq:ff1}):
\begin{equation}
V^{pert(i)}(s)= 1 +a_1s+a_2s^2+...+a_is^i+D^{pert(i)}(s)
\label{eq:peri}
\end{equation}
where $i+1$ subtraction constants are needed to make the last
integral on the RHS of this equation converges and
\begin{equation}
DI^{pert(i)}(s)=  \frac{s^{1+i}}{\pi}\int_{4m_\pi^2}^\infty
\frac{ImV^{pert(i)}(z)dz}{z^{1+i}(z-s-i\epsilon)}
\label{eq:Dperti}
\end{equation}
with $ImV^{pert(i)}(z)$ calculated by the $ith$ loop perturbation
scheme.

Similarly to these equations, the corresponding experimental
vector form factor $V^{exp(i)}(s)$ and $DI^{exp(i)}(s)$  can be
constructed using the same subtraction constants
 as in Eq. (\ref{eq:peri}) but with  the imaginary part replaced by $ImV^{exp(i)}(s)$,
calculated using Eq. (\ref{eq:ieu}).

The one-loop ChPT calculation requires 2 subtraction constants.
The first one is given by the Ward Identity, the second one is
proportional to the r.m.s. radius of the pion. In Fig. 1,  the
imaginary part of the one-loop ChPT calculation for the vector
pion form factor is compared with the result of the imaginary part
obtained from the experimental data. It is seen that they differ
very much from each other. One  expects therefore that the
corresponding real parts calculated by dispersion relation should
be quite different from each other.

In Fig.3  the full real part of the one loop amplitude is compared
 with that obtained from experiment.
\begin{center}
\begin{figure}
  \includegraphics[width=10cm]{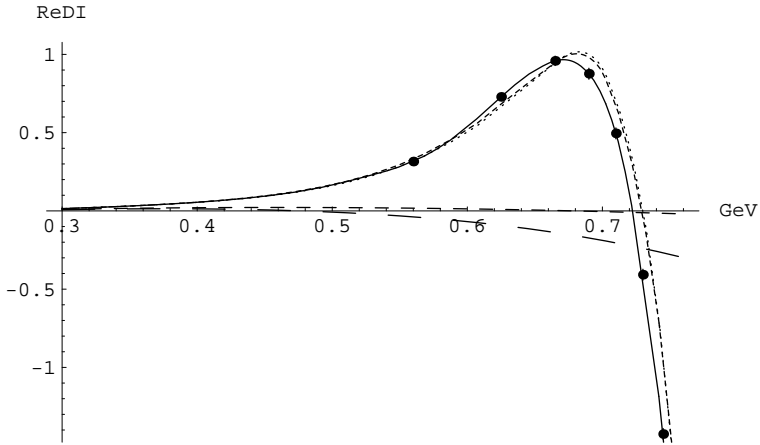}\\
  \caption{Real Parts of the Dispersion Integral ReDI as a function of energy. The label
   of the curves are just the same as in Fig. 1}\label{fig:ReD}
\end{figure}
\end{center}

  At very low energy
one cannot distinguish the perturbative result from the
experimental one due to the dominance of the subtraction
constants. At an energy around $0.56 GeV$ there is a definite
difference between the perturbative result and the experimental
data.  This difference becomes much clearer in Fig. 3 where only
the real part of the perturbative DI, $ReDI^{pert(1)}(s)$,  is
compared with the corresponding experimental quantity,
 $ReDI^{exp(1)}(s)$.
 It is seen that even at 0.5 GeV the discrepancy is clear. Supporters of ChPT would argue
that ChPT would not be expected to work at this energy. One would
have to go to a  lower energy where the data became very
inaccurate.

This  argument is false as can be seen by comparing  the ratio
$R_1=ReDI^{pert(1)}/ReDI^{exp(1)}$ i.e.the ratio of the  one loop
ChPT
  result to its value calculated using the experimental result. We
  have: $R_1=0.16$ for $0\leq s \leq 4m_{\pi}^2$, i.e. the
  one-loop
  ChPT calculated term is too small by a  factor of 7 as compared
  with that calculated using experimental data for the imaginary part
  Above the two pion threshold and up to 600 MeV this value
  becomes even less.

  This ratio becomes better with the inclusion of the two loop
  effects; instead of being  less by a factor of 7 as for the one loop
  calculation it becomes a factor of 2.5 below the two pion threshold
  much larger and becomes larger above the threshold.

  These results illustrate the "amusing" theorem on the small
  analytic function discussed in the introduction.

 Similarly to the one-loop calculation, the  two-loop results are plotted in Fig. (1) - Fig.
(3) \cite{gasser3}.  Although the two-loop result is better than
the one-loop calculation, because more parameters are introduced,
calculating  higher loop effects will not  explain the data.

It is seen that perturbation theory is inadequate for the vector
pion form factor even at very low momentum transfer. This fact is
due to the very large value of the pion r.m.s. radius or a very
low value of the $\rho$ mass $s_\rho$ (see below). In order that
the perturbation theory to be valid the calculated term by ChPT
should be much larger than the
 non perturbative effect. At one loop, by requiring the perturbative calculation
dominates over the non-perturbative effects at low energy,  one
has $s_\rho >>\sqrt{960}\pi f_\pi m_\pi =1.3 GeV^2$ which is far
from being satisfied by the physical value of the  $\rho$ mass.

The unitarized models are now examined. It has been shown a long
time ago that to take into account of the unitarity relation, it
is better to use the inverse amplitude $1/V(s)$ or the Pade
approximant method \cite{truong1, truong3}.

The first model is obtained by introducing a zero in the
calculated  form factor in the ref. \cite{truong1, truong9}  to
get an agreement with the experimental r.m.s. radius . The pion
form factor is now multiplied by ${1+\alpha s/s_\rho}$ where
$s_\rho$ is the $\rho$ mass squared \cite{truong1}: A more insight
to the existence of the zero is possibly that the unitarity
relation was truncated with the two particle state, or the elastic
approximation. The solution of the Muskelishvilli Omnes integral
equation with the inelastic contribution was previously studied
\cite{truong8, truong9} . Its solution can be written as the
product form of the standard form of the elastic unitarity, i.e.
the Omnes function, and the inelastic contribution \cite{truong8}.
Below the inelastic threshold, the inelastic contribution can be
written as a power series whose leading term is the factor
${1+\alpha s/s_\rho}$. This phenomenological description of the
pion form factor with the inelastic contribution was first given
in the reference \cite{truong9}.

The experimental data can be  fitted with a $\rho$ mass equal to
$0.773 GeV$ and $\alpha=0.14$. These results are in excellent
agreement with the data \cite{aleph,na7}.

The second model, which  is more complete, at the expense of
introducing  more parameters, is  based on the two-loop ChPT
calculation with unitarity taken into account. It has the
singularity associated with the two loop graphs. Using the same
inverse amplitude method as was done with the one-loop amplitude,
but  generalizing this method to two-loop calculation, Hannah has
recently obtained a remarkable fit to the pion form factor in the
time-like and space-like regions. His result is equivalent to  the
(0,2)  Pad{\'e} approximant method as applied to the two-loop ChPT
calculation \cite{han}.
  Both models contain ghosts which can be shown to be unimportant
\cite{han}. At this moment there seems to be no preference for one
of either two models, but it would be surprising that the
inelastic effect can completely be neglected in the dispersive
approach.

It is interesting for a given lagrangian and one or two low energy
experimental parameters how do we know whether perturbation theory
was valid. There is of course no general answer to this question.
But by looking at the expression for the one loop perturbation
result and its unitarized version of the pion form factor
\cite{truong1}, one can realize that the r.m.s. radius of the
pion, related to the inverse of the $\rho$ mass, is far too large
to make the perturbation theory valid. This question was discussed
in some details in the reference \cite{truong10}.

Another interesting question is how to incorporate the ChPT
calculation with the vector meson $\rho$  dominance for the pion
form factor which is needed to analyze the experimental data for
the hadronic $\tau$ decays for example. We simply cannot add the
ChPT result to the expression of the vector meson contribution
even with care not to violate the "low energy theorem" of ChPT.
This is so because this procedure would amount to a double
counting. Furthermore, in the time like region, although a rough
fit to the form factor can be made due to the dominance of the
vector meson dominance term, in the  space-like region both the
ChPT term and the vector meson dominance term are roughly equal in
magnitude at moderate momentum transfer and their sum would give a
wrong prediction of the form factor  at moderate momentum
transfer.

It is usually done in ChPT is to do some matching at a few points
of the vector meson dominance term with ChPT the one or two-loop
calculation. Here we encounter the "amusing" theorem and the
inaccuracy of ChPT calculation.

To the best of our knowledge, only the unitarized approach can
preserve the low energy theorem and at the same time gives
correctly the vector meson dominance as required by data,i.e. can
avoid the double counting problem.

 As can be seen from Figs.
1, 2 and 3 the imaginary and real parts of these two models are
very much in agreement with the data. A small deviation of
$ImV(s)$ above $0.9 GeV$ is due to a small deviation of the phases
of $V(s)$ in these two models from the data of the P-wave $\pi\pi$
phase shifts.

In conclusion, higher loop perturbative calculations
 do not solve the unitarity problem. The perturbative
scheme has to be supplemented by the well-known unitarisation
schemes such as the inverse amplitude, N/D and  Pad{\'e}
approximant methods as discussed in the preceeding sections.

\section {$K_{e4}$ Decay}

Although this work was done a long time ago \cite{truong12}, using
the current algebra technique,  reduction formula, dispersion
relation and the equal time current algebra commutation relations,
it is still the only work where the phases and magnitudes and
slope parameters of the S and P-waves form factors can be
calculated and agree with experiments. Because of the lack of the
ChPT two loop calculation for these amplitudes, as pointed out in
\cite{truong1}, the question of calculating the phases of the
relevant form factors were ignored in ChPT calculations.

It is useful to summarize briefly the calculation technique, which
was due originally to Weinberg where the unitarity correction was
neglected \cite{weinberg5}. After extracting the equal time
commutation relation (ETCR) terms, there remains terms which are
proportional to the pion four-momenta which tend to zero in the
soft pion limit; the ETCR terms do not tend to zero and therefore
becomes the low energy theorems and are simply the Born terms. The
terms proportional to the pion momenta, can be shown to obey a
dispersion relation. An integral equation of the
Muskelishvilli-Omnes type can be written with the Born terms(ETCR)
to take into account of the elastic unitarity condition for the
pion pion rescattering. The scale of the relevant amplitudes are
set by the ETCR are in the unphysical region of the $Ke_4$ decay
(the limit of the pion 4-momenta vanishes.

Using the experimental knowledge of the $\pi \pi $ S and P-wave
phase shifts, in the solution of the Muskelishvilli-Omnes integral
equation, the magnitude and phase of the  form factors can be
calculated and agree with experiments.

Because the scale of the problem is set by the low energy current
algebra theorem which is below the two pion threshold, and the
measurement is done at above the two pion threshold, the analytic
continuation of the current algebra result has to pass through the
two pion branch point giving rise to the effect of the square root
threshold singularity \cite{truong12} which is unfortunately
called, nowadays, as the logarithm singularity.

Such an effect enhances the current algebra S-wave result by a
factor of 1.4 (in amplitude) and by a factor of 1.2 for the P-wave
form factor near the two pion threshold in the physical region.
This same enhancement factor is also found in the low energy $\pi
\pi$ scattering and a larger enhancement factor in $\eta
\rightarrow 3\pi$ problem due to the 3 pion final states.

\section{$\eta \rightarrow  3\pi $ Decay}
 Weinberg \cite{weinberg6}, using tree Lagrangian and Dashen theorem :
 \cite{dashen} to calculate the
 rate for $\eta \rightarrow \pi^+ \pi^- \pi^0 $ rate and found its
 width to be 65 eV as compared to the present experimental rate of 295 eV.
 At one loop level the ChPT calculation by  Gasser and Leutwyler
 to be 160 eV . Recent calculations this value is
 increased to 220 $\pm20$ eV.\cite{leutwyler2, bijnens,
 colangelo3}. There are however, difficulties with odd pion
 slope calculation.

 Our approach to this problem was done a long time ago
 \cite{roiesnel} using the integral equation of the type Khuri
 Treiman and Sawyer Wali \cite{khuri}taking into account only of the
 S-wave pion pion interaction. Due to the poor approximation made
 for the treatment of the multiple pion pion scattering a width of
 430 eV was obtained.

 The result of the $\eta \rightarrow  3\pi $ calculation is the same as that of
 $ K \rightarrow  3\pi $ Decay \cite{truong10, truong11} to be discussed
 later, where both the I=0 S-wave  and the I=1 P-wave pion pion
 interactions are taken into account( the P-wave pion pion
 interaction of the pair $\pi^+ \pi^0$ and $\pi^- \pi^0$ are
 allowed but one has to symmetrize  the amplitudes; their contribution only
 affects the odd pion slope and was neglected in the Khuri-Treiman
 integral equation). This calculation yields a width for the
 charged pion mode of $\eta$ to be 240 eV and a correct linear slope
of the odd pion was obtained which was not possible with the ChPT
calculation. With the $\eta$ and $ \eta^{'}$ mixing, the value of
the width is increased to 350 eV.

\section {$K \rightarrow \pi$, $K \rightarrow 2\pi$, $K
\rightarrow 3\pi$ Amplitudes}
 A precise knowledge of the $K\pi$ amplitude and its relation
with $K \rightarrow 2\pi$ and $K \rightarrow 3\pi$ are of a
fundamental importance in the study of the origin of the $\Delta
I=1/2$ rule, the CP violation effect in the standard model and the
rare decays of $K_L$ and $K_S$.

The simplest way to implement chiral symmetry and SU(3) on $K
\rightarrow 2\pi$ and $K \rightarrow 3\pi$ is to use the chiral
lagrangian \cite{cronin}. The amplitude for $ M(K_S \rightarrow
\pi^+\pi^-$) is given by:
\begin{equation}
M(K_S(k) \rightarrow \pi^+(p)\pi^-(q)) =\frac{1}{2}iCf_\pi
(2k^2-p^2-q^2) \label{eq:k2pi}
\end{equation}
The amplitude for $M(K_L \rightarrow \pi^+\pi^-\pi^0)$ is
\begin{equation}
M(K_L \rightarrow
\pi^+(q_1)\pi^-(q_2)\pi^0(q_3))=\frac{C}{\sqrt{2}}(s-\mu^2)
\label{eq:k3pi}
\end{equation}

and the $K-\pi$ amplitude is:

\begin{equation}
M(K_L \rightarrow \pi^0) = -\frac{C}{\sqrt{2}}f_\pi^2q(\pi)q(K)
\label{eq:kpi}
\end{equation}
 where $s=(q_1+q_2)^2, t=(q_1+q_3)^2, u=(q_2+q_3)^2$ with
$s+t+u= 3s_0=3\mu^2+m^2$ where m and $\mu$ are respectively K and
$\pi$ masses.

Using the experimental determination of the $\Delta=1/2$ amplitude
$a_1/2=(0.469\pm0.006).10^-3 MeV$ it is found that
$C=1.26.10^-11MeV$. The $K\rightarrow 3\pi$ as given by Eq.
(\ref{eq:k3pi})is:
\begin{equation}
M(K_L \rightarrow
\pi^+\pi^-\pi^0)=7.43.10^-7(1+0.233(s-s_0)/\mu^2) \label{eq:kl3}
\end{equation}
This result is to be compared with the experimental value:
\begin{equation}
M(K_L \rightarrow
\pi^+\pi^-\pi^0)=9.10.10^-7(1+0.264(s-s_0)/\mu^2) \label{eq:kl3ex}
\end{equation}

It is seen that the tree lagrangian yields a prediction for the
$K_L \rightarrow 3\pi$ amplitude a value too low by 20\% and the
odd pion slope is 12\% too low. Such discrepancies are due to the
neglect of unitarity. For example the Eq.(\ref{eq:k2pi})is purely
real while unitarity requires it to have a phase of approximately
of $40^0$ onthe K mass. The resolution of these problems was given
 by taking unitarity into account for the
rescattering in the S-wave as well as P-wave pion pion
interactions \cite{truong10}. It should be remarked that due to
the 3-body in the final state, we can have both S and P-waves pion
pion interaction.
\begin{equation}
M(K_L \rightarrow
\pi^+\pi^-\pi^0)=8.86.10^-7(1+0.250(s-s_0)/\mu^2)
\label{eq:kl3truong}
\end{equation}
and is very much in agreement with the experimental result
Eq.(\ref{eq:kl3ex})

\section {Study of
$\gamma \gamma \rightarrow 2\pi$, $K_S \rightarrow 2\gamma$ and
$K_L \rightarrow \pi^0\gamma\gamma $}
 The result of the calculation of the $K \rightarrow \pi$
 calculation discussed in the previous section allowed us to
 calculate the rare decay modes of $K_S \rightarrow 2\gamma$ and
$K_L \rightarrow \pi^0\gamma\gamma $. These were done in the
reference \cite{truong11} and there are agreements between theory
and experiments and will not be discussed here. I would like to
point out two old papers on $\gamma \gamma \rightarrow 2\pi$ which
are still relevant for further study on this subject \cite{rosner,
budnev}. A more recent paper is also relevant\cite{donoghue}. 

\section{$\tau \rightarrow K \pi \nu$ and $\tau \rightarrow 3\pi \nu$
Decays}

The process $\tau \rightarrow K \pi \nu$ illustrate the usefulness
of the combination of the current algebra low energy theorems with
unitarity and dispersion relation \cite{beldjoudi1}. The S and
P-wave $K-\pi$ form factor were calculated using elastic unitarity
in combination with dispersion relation. The KSRF relation were
found to be valid, and the forward-backward asymmetry due to the
interference of the S and P wave form factors were predicted and
to be verified by future experimental results.

A similar calculation was done for the $\tau \rightarrow 3\pi \nu$
decay \cite{beldjoudi2}.

\section{Enhancement Factor in $K \rightarrow 2\pi$}
Discussions in the previous sections are based on Current Algebra
or Effective Lagrangian low energy theorems. Given these theorems
at low energy, usually in the unphysical region, with elastic
unitarity and dispersion relations we analytically continue these
theorems to the elastic region. This is a small extrapolation. In
a different line of physics, we want to ask a much more difficult
question what is the difference in amplitudes, e.g. in the $K
\rightarrow 2\pi$ with the final state interaction of the two pion
interaction when it is switched on and off. This last  question
has no answer and is dependent on the input assumption.

This question was asked a long time ago by Isgur et al
\cite{isgur}. The answer they gave was that the calculation of the
matrix element with the final state interaction switched off has
to be multiplied with the S-wave pion pion interaction wave
function at the origin. This question was examined
\cite{truong30}, and the answer depends on what one assumes when
writing down the Muskelshivilli-Omnes integral equation. Namely,
one assumes that at infinite energy the pi pi interaction does not
affect the result of calculation, in other words, the pi pi
interaction are completely switched off. and that the elastic
unitarity is valid for writing down the dispersion relation for
the imaginary part of the enhancement factor. Under these
assumptions, the enhancement factor is simply the inverse of the
Jost function. The Jost function has the property that it becomes
unity at infinite energy which is an extremely long extrapolation
from the elastic pi pi region to an infinite energy. The answer
given needs not be reliable. For a detailed discussion on this
problem, the reader is referred to the reference \cite{truong30}.


\section {Acknowledgement}
I would like to thank Prof. Hoang Ngoc Long and the organizing
committee for  inviting me to this conference, and for giving me a
chance to present this talk before my definite retirement from
physics. I would like also to thank them to give me a chance to
express my gratitude to my thesis advisor, H.A. Bethe, and also to
F.J.Dyson, T.D. Lee, K. Nishijima and C.N. Yang who have taught me
physics. They are, however, not responsible for errors that I made
in this article.

\end{document}